\documentclass[a4paper,12pt]{amsart}
\usepackage[margin=3cm]{geometry}
\usepackage{graphicx}
\usepackage{url}
\usepackage{color}

\usepackage{setspace}

\begin{document}

\title[Normal phylogenetic networks]{``Normal'' phylogenetic networks may be emerging as the leading class.}

\author{Andrew Francis} 
\address{School of Mathematics and Statistics, University of New South Wales, Sydney, Australia, a.francis@unsw.edu.au.}

\begin{abstract}
The rich and varied ways that genetic material can be passed between species has motivated extensive research into the theory of phylogenetic networks.  Features that align with biological processes, or with desirable mathematical properties, have been used to define classes and prove results, with the goal of developing the theoretical foundations for network reconstruction methods.  We may have now reached the point where a collection of recent results can be drawn together to make one class of network, the \emph{normal} networks, a leading contender, sitting in the sweet spot between biological relevance and mathematical tractability.  

\end{abstract}
 \maketitle

The need to be able to represent evolutionary relationships that cannot be shown by familiar phylogenetic trees has been recognised for half a century now~\cite{sneath1975cladistic,hilario1993horizontal,doolittle1999phylogenetic}.  These relationships, such as horizontal gene transfer (HGT), endosymbiosis, and hybridization, play a significant evolutionary role, and so how to infer and represent them, using phylogenetic networks, is increasingly important~\cite{DeSalle2020should,doolittle2007pattern-plurali,dagan2006the-tree-of-one}.  How exactly to {do} that has been described as one of the key problems in the use of mathematics in biology~\cite{cohen2004mathematics}.  

The last couple of decades has seen significant investment in exploring the modelling, inference, and mathematics of phylogenetic networks.  
The difficulty is that, in contrast to phylogenetic trees, there are infinitely many phylogenetic networks that relate a given set of taxa appearing at the leaves of the network.  This makes inference decidedly harder: it is not even \emph{theoretically} possible to search the whole space for optimal solutions.

In order to infer phylogenetic networks, researchers have focused on classes of networks that have important provable properties.
To this end, a menagerie of classes has been proposed, each capturing features whose goal is either to represent a specific biological process, such as horizontal gene transfer, or to provide properties that are amenable to mathematical persuasion.  Some of these latter classes include tree-child networks~\cite{cardona2009comparison}, tree-based networks~\cite{francis2015which-phylogene}, orchard networks~\cite{erdos2019class}, regular networks~\cite{willson2011regular}, and reticulation-visible networks~\cite{Bordewich2016reticulation}.  

In this article, we do not attempt a general survey of research in phylogenetic networks; see~\cite{kong2022classes} for a recent such survey.  Rather, we attempt to put forward the case that one of these classes --- the \emph{normal} networks~\cite{willson2010properties-of-n} --- is emerging as the strong contender for inference, because of its proven mathematical properties and prospects for practical use, that we describe below.  It is now reasonable to hope that further research on normal networks might bring practical inference of phylogenetic networks within reach.

\section*{The ecosystem of phylogenetic networks}

Phylogenetic networks are thought of in two broad paradigms: implicit networks, and explicit networks.  Implicit networks are ways to represent the relationships among species, usually inferred from a distance metric on the leaves, and are ``unrooted'', which generally means that there is no direction on the edges.  Examples are those obtained from software such as SplitsTree~\cite{huson2006application} or Network 10~\cite{network10}.  They are called implicit because the internal vertices are not intended to represent actual historical evolutionary events, but rather, are effective ways to summarize the genetic data.  In contrast, {explicit} networks are rooted, and directed, and non-leaf vertices are meant to represent possible events such as speciation, or reticulation (hybridization or horizontal gene transfer).  This paper focusses on explicit networks, which we define as follows.

\begin{figure}[ht]
\begin{center}
\includegraphics{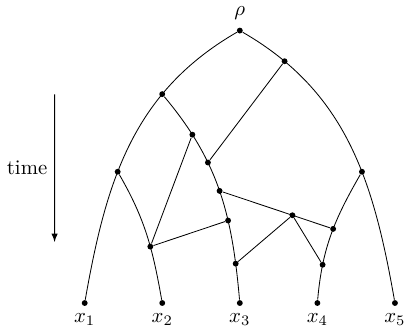}
\caption{A phylogenetic network. Edges are directed forward in time (down the page, arrows on edges omitted), from the root $\rho$ to the leaves $x_i$. }
\label{f:network.eg}    
\end{center}
\end{figure}

A \emph{phylogenetic network} is a directed acyclic graph with vertices of the following types: \emph{the root} (in-degree 0 and out-degree $\ge 1$); \emph{tree vertices} (in-degree 1 and out-degree $\ge 2$); \emph{reticulation vertices} (in-degree $\ge 2$ and out-degree 1); and \emph{leaves} (in-degree 1 and out-degree 0).  See Figure~\ref{f:network.eg} for an example.  If the internal (non-leaf and non-root) vertices all have total degree 3, and the root has out-degree 2, the network is referred to as \emph{binary}.

Many families of phylogenetic network have been defined to date, each seeking to capture either biological intuition or desirable mathematical features.
``Normal'' networks are one such class, and are a subclass of the ``tree-child'' networks.  They were first defined, and given that name, by Willson in 2010~\cite{willson2010properties-of-n}.  

\begin{figure}[ht]
\begin{center}
\includegraphics{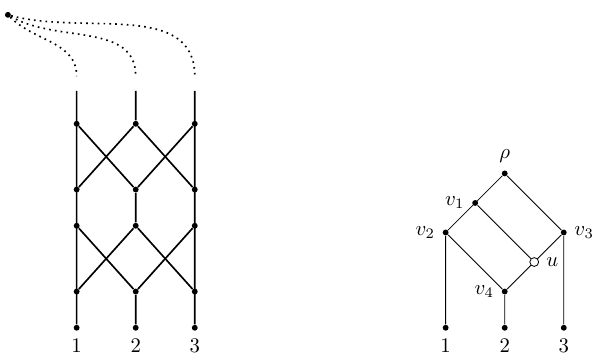}
\caption{Left: Phylogenetic networks can be unbounded in the number of vertices, for any fixed number of leaves.  Right: A network with a vertex $u$ that is not visible, shown as an unfilled dot. For each leaf, there is at least one path from the root to that leaf that avoids $u$, for instance $\rho\to v_1\to v_2\to 1$; $\rho\to v_1\to v_2\to v_4\to 2$; and $\rho\to v_3\to 3$.  The vertices $v_i$ are all visible, as for each there is a leaf for which all paths from the root to the leaf pass through it. For instance, all paths going from $\rho$ to leaf 1 go through $v_1$ and $v_2$, so they are both visible.}
\label{f:infinite}    
\end{center}
\end{figure}

A network is \emph{tree-child} if every internal vertex in the network has a child that is a tree vertex or leaf~\cite{cardona2009comparison}. Each vertex $v$ in a tree-child network has the property that there is a leaf $x$ for which every path from the root to $x$ passes through $v$~\cite[Lemma 2]{cardona2009comparison}.  This property is called ``visibility'' because it implies that the evolutionary signal from $v$ is visible in a leaf, $x$ (see Figure~\ref{f:infinite} for an example).  

A \emph{normal network} is simply a tree-child network without short-cuts (edges $(u,v)$ for which there is another path $u\to v$ in the network)~\cite{willson2010properties-of-n}.  

It is worth emphasizing the significance of the property that all vertices are visible in a normal (or tree-child) network.  Visibility is a mathematical property about the underlying graph of the network, but it carries significant power in terms of its biological meaning, and in particular what it means for inference from biological data at the leaves.  The key point is that if a vertex is not \emph{necessarily} on a path from the root to any particular leaf, then the signal from that vertex may be completely absent from the information in the leaves.  That makes the case for inferring the existence of that vertex very weak.  We will refer again to visibility in later sections, because of its centrality.

The class of tree-child networks is in turn contained in the classes of \emph{orchard} networks~\cite{erdos2019class} and \emph{tree-sibling} networks~\cite{cardona2008distance}, which are briefly defined here.  The orchard networks are those that can be progressively reduced via ``cherry reductions'' and ``reticulated cherry reductions'' at the leaves, to a single vertex.  They have the attractive property of being characterised by their ``ancestral profiles'', which are a set of tuples for the leaves of the network that enumerate the number of paths from each vertex to each leaf~\cite{erdos2019class}. 
Tree-sibling networks are those for which every reticulation vertex has a tree vertex as a sibling (a vertex that shares the same parent).   Both the orchard and tree-sibling  classes in turn are contained in the class of \emph{tree-based} networks, which are networks that have a spanning tree whose leaves are those of the networks~\cite{francis2015which-phylogene}, and the class of \emph{labellable} networks, which are those whose internal vertices can be deterministically labelled~\cite{francis2023labellable} (many classes, including normal networks, can be characterised in the labellable framework~\cite{francis2024phylogenetic}).

\begin{figure}[ht]
\begin{center}
\includegraphics{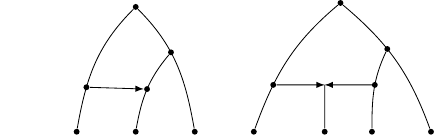}
\caption{Phylogenetic network representations of a horizontal gene transfer (HGT) on the left, and a hybridization, on the right.}
\label{f:hybrid.HGT}
\end{center}
\end{figure}

Networks that model only specific processes, such as HGT or hybridization, can be formally defined as trees with additional types of event permitted, and these constraints provide more specific restrictions that help narrow the focus within this ecosystem of classes.  \emph{HGT networks} and \emph{hybridization networks}, represent the named event as an almost instantaneous (horizontal) effect on the network, as shown in Fig.~\ref{f:hybrid.HGT} (for example, see~\cite{gogarten1995early}). In contrast, other families of network usually represent reticulate events with ambiguous timing, which allows uncertainty in the representation (that is, reticulate arrows need not be horizontal), as shown in the example in Fig.~\ref{f:network.eg}.  

\begin{figure}[]
\begin{center}
\includegraphics[width=\textwidth]{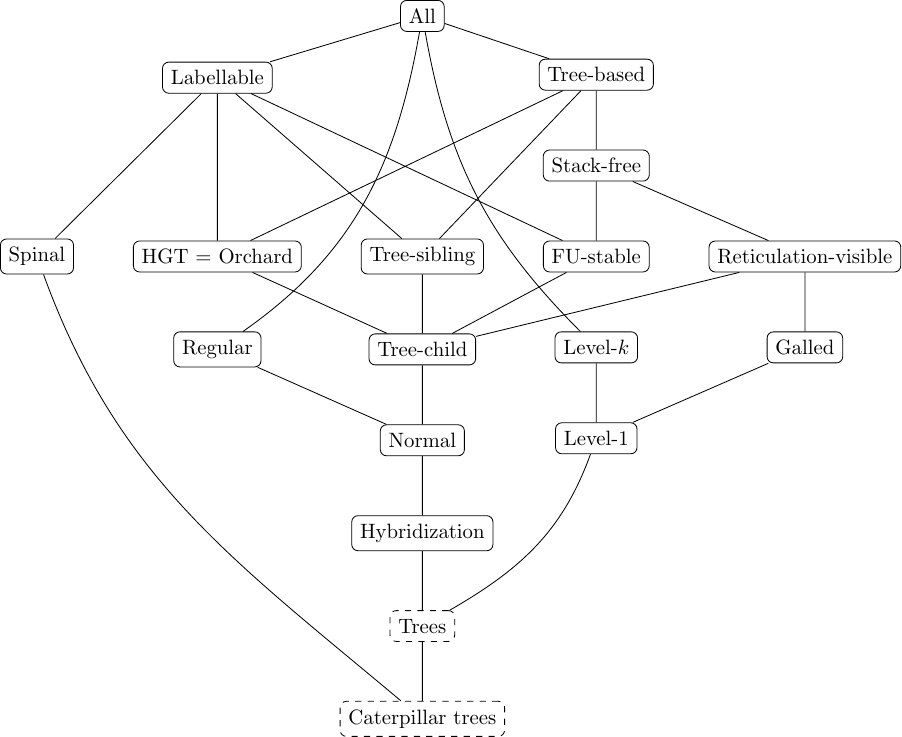}
\caption{Inclusion relationships among some classes of rooted binary phylogenetic networks, where edges denote inclusion of a class in the class above. All class names are classes of networks, including the special classes of Trees and Caterpillar trees shown in dashed boxes at the bottom. This figure is adapted from~\cite[Fig 12]{kong2022classes}, showing a subset of networks from that figure while including some additional classes and newly known relationships.  The additional classes shown are the recently-defined labellable and spinal network classes~\cite{francis2023labellable,francis2024phylogenetic}, and the hybridization class. To connect spinal networks with trees, Caterpillar trees have been added below, as a subclass of Trees.  The new relationships are that hybridization networks are all normal~\cite[Prop 10.12]{steel2016phylogeny}, that the HGT and orchard network classes are identical~\cite{van2022orchard}, and that fold-unfold (FU)-stable networks are labellable (this follows from~\cite[Thm 1]{huber2016folding} and the characterization of labellable networks~\cite[Thm 3.3]{francis2023labellable}).  Note that stack-free and reticulation-visible networks are not labellable, because there are many such networks that violate~\cite[Thm 3.3]{francis2023labellable}. For definitions and references for the other relationships in the figure, see~\cite{kong2022classes}. }
\label{f:networks}
\end{center}
\end{figure}

Results relating the many classes of phylogenetic network regularly appear.  A thorough recent summary of many of them and their properties is given in~\cite{kong2022classes}.  
Relationships among some of the key families of phylogenetic networks mentioned in the present paper are shown in Fig.~\ref{f:networks}.

For instance, it turns out that \emph{binary} hybridization networks are normal networks~\cite[Prop.~10.12]{steel2016phylogeny}.  If one is seeking a network that describes the evolutionary history of a set of eukaryotic species, this is the sort of network that one is most likely to want to reconstruct.  If on the other hand, one is seeking a network describing prokaryotic evolution, or even tumor evolution (see e.g~\cite{valcz2022small}), then a horizontal gene transfer (HGT) network might be more relevant.  HGT networks allow more (mathematical) freedom, because each event only requires one arc.  Despite this, they have recently been shown to be precisely the orchard networks~\cite{van2022orchard}, which contains the class of normal networks.  

In the rest of this paper we lay out three significant features that --- on top of those already mentioned --- support the possibility that the class of normal phylogenetic networks is where we are most likely to see the cross-over from theory into practical inference.

\section*{Case for normal networks I: Reconstruction}

A central application of the research into phylogenetic networks is to devise methods that enable the \emph{reconstruction} of a reticulate evolutionary history, using the data available in the present.  It is possible to reconstruct a phylogenetic {tree} from the set of ``rooted triples'' on the leaves (the rooted triples in a tree are the restrictions of the tree to sets of three leaves)~\cite{aho1981inferring}, and tree reconstruction is used every day in diverse fields (for instance epidemiology, anthropology, linguistics, taxonomics). But self-evidently, trees cannot capture the many reticulate processes that are known to occur, and that leave a signal in the data at the leaves. Network reconstruction is significantly more difficult than that for trees, for several reasons.  

The first is that the space of networks for a given number of leaves is infinite.  Once speciation is not the only event, and once branches can merge, an intertwined set of edges can potentially stretch back from the leaves without bound (see Fig.~\ref{f:infinite}).  

The second is that if a reticulation occurred far enough back in time, and if there were subsequent other speciation and reticulation events, it can be possible that any signal of that reticulation has not survived to the leaves.  In particular, if there is a path from the root to each leaf that avoids a reticulation vertex, there may be no evidence of that vertex in the leaves at all.  Such a vertex would be ``invisible'' to us in the present, and this is the intuition behind the notion of ``visibility'' mentioned elsewhere.

These two difficulties force us to restrict our ambitions from reconstruction of any network whatsoever, to reconstruction of a network from a restricted class.  The questions then are ``which class?'', and ``what information is sufficent to reconstruct a network from it?''.  The case for normal networks (to answer the first question) in the context of reconstruction is precisely that it has been possible to prove powerful results that answer the second.

While trees can be reconstructed from their rooted triples (the relationships among sets of three leaves), an arbitrary network cannot be reconstructed from its {displayed} trees, even if one has the \emph{full set} of trees displayed by the network~\cite{pardi2015reconstructible}. 
However, for normal networks there are results giving reconstruction from either substructures, or from the set of trees displayed by the network.

Firstly, like trees, normal networks can also be reconstructed from more simple substructures: the sets of triplets, and four-leaf ``caterpillar trees'' (caterpillars are a simple tree-structure that have just a single cherry: a pair of leaves with the same parent)~\cite{linz2020caterpillars}.  Note the similarity to trees, whose reconstruction requires just the set of their rooted triples.  To reconstruct a normal network, all one needs is what one needs for trees (relationships among the subsets of three taxa), and in addition, the relationships among some specific subsets of four taxa.

At a larger scale than substructures on three or four leaves, the trees displayed by a network (which have the same leaf-set as the network), along with their multiplicities, are important, because they are often interpreted as representing the evolutionary histories of individual genes within the population. For normal networks, early results showed that in the binary case (in which each internal vertex has degree 3), they are among those that \emph{can} be reconstructed from their displayed trees~\cite{willson2011regular}.  But there are stronger facts about displayed trees for normal networks.  For instance, normal networks display any given tree at most once~\cite{willson2012tree-average-di,cordue2014phylogenetic-ne}.  This means that for binary normal networks with $r$ reticulations, there are exactly $2^r$ trees displayed in the network.  Most powerfully, a normal phylogenetic network is \emph{uniquely determined} by these $2^r$ trees that it displays, and indeed can be reconstructued from those trees by a polynomial time algorithm~\cite{bordewich2024set}.

Other classes of network also have some reconstructability results, but these tend to be more limited or conditional, and typically require sub\emph{network} information (rather than sub\emph{tree}).  For instance, networks in the larger class of tree-child networks can be reconstructed from their displayed trees, but not uniquely~\cite{zhang2023fast,zhang2024phylofusion}. And for reconstruction from sub\emph{networks}, binary orchard networks (and therefore also tree-child and normal networks) are encoded by their ``trinets'', and can be reconstructed in polynomial time~\cite{semple2021trinets} (trinets are rooted phylogenetic networks on three leaves~\cite{huber2013encoding}),  generalising a similar result for binary tree-child networks~\cite{van2014trinets}.

\section*{Case for normal networks II: Identifiability}

Another side of the reconstruction problem, and the second element of the case for normal networks, is the property of \emph{identifiability}.  Identifiability is a property of a statistical model: a model is identifiable if its parameters can be uniquely determined from the probability distribution generated under the model.  In the context of phylogenetic networks, the model is a process of evolution in time that begins with a single (root) taxon (or genome), and produces a set of taxa as output.  The network on which evolution is occurring is, itself, one of the key parameters of such an evolutionary model.  Putting the networks to the fore, a class of phylogenetic networks is identifiable for a particular evolutionary model if, given the distribution of outputs of the model (DNA sequences of taxa at the leaves of the network), it is possible to uniquely determine the network (in that class) on which the sequences arose.

Evolution on a network is generally modelled as a generalisation of the standard model for evolution on a tree (see~\cite{allman2006identifiability} for an introduction).  That is, for each site on the genome, there is a fixed probability of a mutation to another nucleotide along an edge of the network.  The process moves along the sequence, one nucleotide at a time from the root evolving down edges of the tree to the leaves.

There are two ways that a reticulation can be handled by this model.  One widely-studied model treats each reticulation as a choice between parents, randomly choosing which one will continue along the edge below the reticulation~\cite{nakhleh2010evolutionary}.  This independence between nucleotides makes the model amenable to a range of interesting methods from algebraic geometry, but so far identifiability has been challenging to prove beyond Level-1 networks~\cite{gross2018distinguishing,hollering2021identifiability}.  

Another model, adapted from one for pedigrees~\cite{thatte2013reconstructing}, instead chooses an in-bound edge for each reticulation vertex, and evolves the site on the resulting tree.  As the process moves along the alignment, there is a (low) probability of changing the in-bound edge at each reticulation, and hence changing the tree the sequence is evolving on.  As a consequence, blocks of the resulting sequence evolve under a common tree that is displayed by the network.  It could be argued that this more closely models reticulate processes, where for instance horizontal gene transfer involves blocks of a sequence being inserted from elsewhere, reflecting evolution under a different tree.  The additional structure resulting from this model has allowed stronger results to be proven, in particular, that \emph{all binary normal networks are identifiable} under it~\cite{francis2018identifiability}.  Indeed, a slightly larger class has been shown to be identifiable: all binary tree-child networks except those with a short-cut at the root.

\section*{Case for normal networks III: Mathematical workability and universality}

The third part of the case is that normal networks have a set of mathematical properties that are unique, and biologically informative.  

For instance, {every} internal tree vertex in a normal network is the least common ancestor (in the graph theoretic sense) of two leaves~\cite{willson2010properties-of-n}.  This means that every internal tree vertex carries a specific biological piece of information, and is a complement to the fact that every vertex, whether tree or reticulation, is visible in a normal network (as mentioned earlier).  

Some other mathematical properties have already been described in previous sections on reconstruction and identifiability, for instance: a normal network with $r$ reticulations displays exactly $2^r$ distinct trees, is uniquely determined by the trees it displays, and can be reconstructed from those trees by a polynomial time algorithm.  The broader question of whether a given set of trees can be displayed by a normal phylogenetic network, can also now be determined algorithmically, using ``cherry-picking'' sequences~\cite{bordewich2024set}.

These results are surprisingly strong, and point to a significant efficiency in the class of normal networks: normal networks represent information that we have, but with minimal redundancy.

Finally, the icing on the cake.  There is a sense in which focussing on normal networks is not in fact neglecting other classes of network, for two significant reasons.

The first is that it has recently been shown that in the limit, the normal networks are effectively \emph{all} phylogenetic networks, as long as the number of reticulations is bounded relative to the number of leaves~\cite{fuchs2024asymptotic}.  
More precisely, if the number of reticulations is constant, or grows sufficiently slowly with $n$, then the proportion of all possible binary phylogenetic networks on $n$ leaves that are normal tends to 1 as $n$ grows.

The second recent development is the result that {every} phylogenetic network has a uniquely defined ``normalization''.  That is, there is a well-defined and polynomial time process that discards redundant vertices and edges in a network to yield a canonical normal network~\cite{francis2021normalising}.  The process takes the set of visible vertices from the network, then constructs their Hasse diagram based on directed paths in the original network, and finally suppresses redundant vertices of degree two.

\begin{figure}[ht]
\includegraphics[width=\textwidth]{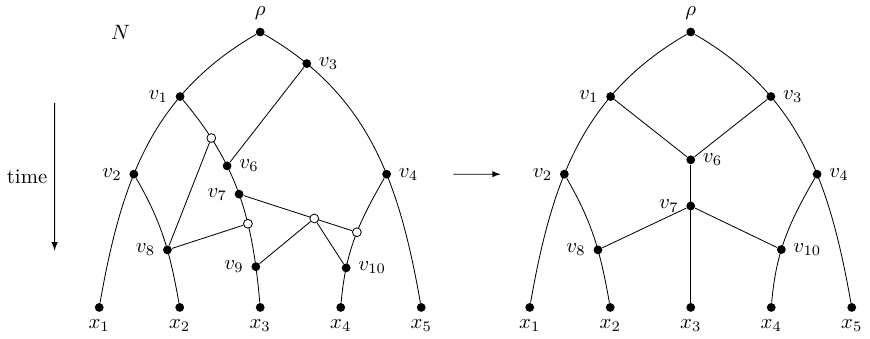}
\caption{A phylogenetic network $N$, from Figure~\ref{f:network.eg}, and its normalisation.  Vertices are shown solid if they are visible.  The non-visible vertices in $N$ are those vertices $v$ with the property that for any leaf $x_i$ there is a path from the root of $N$ to $x_i$ that does not pass through $v$.
Thus, tracing back from any extant leaf to the root, it is possible to never ``see'' a non-visible vertex as there is always a path that avoids it.
The normalisation of $N$ on the right shows all visible vertices in $N$ except $v_9$, which is suppressed since it would be degree 2 in the cover digraph of the visible vertices of $N$. 
}
\label{fig}
\end{figure}

An example of a normalization is shown in Figure~\ref{fig}. The normalization captures the key information in \emph{any} given network: the information that is in fact reconstructable, that is, the vertices and edges whose evolutionary signal may be visible in the leaves.  If one has data that evolved from a process on a network, reconstructing the normalization of the network is the sweet spot: both the most information reasonable to hope for, and the most that can practically be obtained.  

Other ways to take a quotient (a compression) of a network to a simpler one capturing key information, are being developed. For instance, one can form a ``distinct cluster tree-child'' (DCTC) network from any network~\cite{willson2022distinct}.  This new class has some attractive properties worth further investigation, but for the moment these are not comparable with the advantages presented by normal networks. 

\section*{Conclusion}

Phylogenetics is the science of inferring ancestral relationships among a set of present-day species.  With such a project, and in the presence of reticulate events such as hybridization and horizontal gene transfer, there can inevitably be information whose signal is irretrievably lost.  As the mathematical properties of phylogenetic networks continue to be developed, we can expect clarity to emerge about which classes of network are recoverable from different kinds of data, and we can expect the development of robust methods to make inferences.  The current evidence, as summarized here, seems to be indicating that the class of \emph{normal} networks may provide a framework for this emergent clarity, and for inference methods that respect the information constraints imposed by reticulation. 

There will certainly be more nuance to come.  We have not discussed details of several dimensions of network inference, making at most a passing reference to implicit networks, or the related semidirected networks~\cite{gross2021distinguishing,linz2023exploring}, and have not gone far into networks that arise from different types of data or processes (aside from a brief mention of HGT and hybridization networks).
The further development of methods for the reconstruction of normal networks, and results that develop their mathematical properties, seems a most promising direction for future research.

\end{document}